\def\year{2018}\relax
\documentclass[letterpaper]{article} 
\usepackage{aaai18}  
\usepackage{times}  
\usepackage{helvet}  
\usepackage{courier}  
\usepackage{url}  
\usepackage{graphicx}  

\usepackage{amsmath}
\usepackage{amsthm}
\usepackage{amssymb}
\usepackage{subfig}
\usepackage{epstopdf}
\usepackage{wrapfig}
\usepackage[ruled,vlined]{algorithm2e}
\usepackage{multirow}
\usepackage{paralist}

\begin{document}
%
\title{Slim-DP: A Light Communication Data Parallelism for DNN}
\author{Shizhao Sun$^{1,}$\thanks{This work was done when the author was visiting Microsoft Research Asia.}, Wei Chen$^2$, Jiang Bian$^2$, Xiaoguang Liu$^1$ \and Tie-Yan Liu$^2$ \\
	$^1$College of Computer and Control Engineering, Nankai University, Tianjin, 300071, P. R. China\\
	$^2$Microsoft Research, Beijing, 100080, P. R. China \\
	sunshizhao@mail.nankai.edu.cn, wche@microsoft.com, jiabia@microsoft.com,\\
	liuxg@nbjl.nankai.edu.cn, tyliu@microsoft.com
}
\maketitle
\begin{abstract}
	Data parallelism has emerged as a necessary technique to accelerate the training of deep neural networks (DNN). In a typical data parallelism approach, the local workers push the latest updates of all the parameters to the parameter server and pull all merged parameters back periodically. However, with the increasing size of DNN models and the large number of workers in practice, this typical data parallelism cannot achieve satisfactory training acceleration, since it usually suffers from the heavy communication cost due to transferring huge amount of information between workers and the parameter server. In-depth understanding on DNN has revealed that it is usually highly redundant, that deleting a considerable proportion of the parameters will not significantly decline the model performance. This redundancy property exposes a great opportunity to reduce the communication cost by only transferring the information of those significant parameters during the parallel training. However, if we only transfer information of temporally significant parameters of the latest snapshot, we may miss the parameters that are insignificant now but have potential to become significant as the training process goes on. To this end, we design an Explore-Exploit framework to dynamically choose the subset to be communicated, which is comprised of the significant parameters in the latest snapshot together with a random explored set of other parameters. We propose to measure the significance of the parameter by the combination of its magnitude and gradient. Our experimental results demonstrate that our proposed Slim-DP can achieve better training acceleration than standard data parallelism and its communication-efficient version by saving communication time without loss of accuracy.
\end{abstract}

\section{Introduction}\label{sec:intro}
Rapid development of deep neural networks (DNN) has demonstrated that its great success is mainly due to the power of big models learned based on big data~\cite{srivastava2015training,he2015deep}. However, the extremely time-consuming training has become a critical debt to obtain a large-scale DNN model. To accelerate the training of DNN, data parallelism~\cite{dean2012large,zhang2015deep,chen2016revisiting,povey2014parallel,chen2016scalable} has emerged as a widely-used technique in recent years. In a typical data parallelism approach, after distributing all the training data into a set of local workers, the training procedure continues iterations of the three main steps: first, each worker independently trains the local model based on its own local data; then, the learned parameter updates are pushed to the parameter server~\cite{li2014communication,li2014scaling} and aggregated to the global model; finally, the local worker pulls the new snapshot of the global model from the parameter server, and set it as the new starting point for the next training iteration. As it transfers the updates and parameters of the whole model between local workers and the parameter server, we refer this standard data parallelism approach as \emph{Plump Data Parallelism}, abbreviated as \emph{Plump-DP} in the following of this paper.

Although such typical data parallelism framework is well-motivated, it cannot achieve satisfactory training acceleration in practice since it suffers from the heavy communication cost by transferring huge amount of parameters and updates between local workers and the parameter server~\cite{dean2012large,seide2014,alistarh2016qsgd}. For example, as shown in~\cite{alistarh2016qsgd}, for the parallel training of AlexNet~\cite{krizhevsky2012imagenet}, GoogLeNet~\cite{szegedy2015going}, ResNet152~\cite{he2015deep} and VGG-19~\cite{simonyan2014very}, the communication takes 10\% to 50\% of the overall training time when there are 8 local workers, which are non-negligible compared to the training time; and the percentage of communication time continues to grow when the number of local workers is increased.

In-depth understanding on DNN reveals that it is usually highly redundant, in that a considerable proportion of parameters of a well-trained DNN model is insignificant to the model performance. In the model compression task, for example, experiments in~\cite{han2015learning} indicated that we can delete about 90\% of parameters of the well-trained AlexNet and VGG-16 model without significantly influencing the model performance via weight pruning. The redundancy of the DNN model exposes a great opportunity to reduce the communication cost in the parallel training of DNN without the loss of accuracy. We may only need to transfer the computing information of those significant parameters during the parallel training of DNN. For ease of reference, we call the subset of parameters communicated during the parallel training of DNN \emph{communication set}. 


However, it is improper to only communicate the computing information of temporally significant parameters in the current global model during the parallel training of DNN. The reason is that different from removing the redundant parameters of a well-trained model in the model compression task, the selection of significant parameters in the parallel training of DNN is conducted during the training process, which will influence the forthcoming optimization path. If we only communicate the information of the temporally significant parameters, we may ignore the parameters that is insignificant now but have potential to become significant as the training process goes on if their computing information is communicated. 

To this end, we design an Explore-Exploit framework to dynamically identify the communication set, which is comprised of significant parameters in the latest snapshot of the global model together with a random explored set of other parameters, considering that these parameters are worthy of being explored to become significant before the training converges. For sufficient exploration, the random set of parameters is frequently re-sampled for every communication. For better exploitation, the set of significant parameters is reselected for every $q$ rounds of communication because communicating this set of parameters for multiple times can help us to learn the parameters in this set more sufficiently. Due to its objective of substantial reduction of communication cost, we name our proposed framework as \emph{Slim Data Parallelism}, abbreviated as \emph{Slim-DP} in the following of this paper.

To measure the significance of the parameters, we consider two factors: the magnitude of this parameter and the magnitude of its gradient. The small magnitude of the parameter indicates that deleting this parameter will not significantly decline the model performance, which is also widely used in model compression task~\cite{han2015learning}. The magnitude of the gradient of one parameter indicates the sensitivity of the loss function with respect to this parameter~\cite{lecun1990optimal,hassibi1993second}. The large magnitude of the gradient indicates that the deletion of the parameter will cause significant change to the loss function even if the magnitude of the parameter is not such great.

To verify the effectiveness of Slim-DP, we conducted experiments on ImageNet~\cite{ILSVRC15}, and evaluate our method on two models, i.e., GoogLeNet and VGG-16. We have following observations from the experimental results: 1) compared to Plump-DP, Slim-DP can save about 55\% and 70\% of the communication time for GoogLeNet and VGG-16 respectively; 2) by saving communication time, Slim-DP runs about 1.1 and 1.3 times faster than Plump-DP to process the same amount of data for GoogLeNet and VGG-16 respectively; 3) by saving communication time as well as benefiting from ignoring the redundant information, Slim-DP runs about 1.2 and 1.4 times faster than Plump-DP to achieve the same or even better convergence for GoogLeNet and VGG-16 respectively; 4) Slim-DP also outperforms other communication-efficient version of Plump-DP, i.e., Quant-DP, for both GoogLeNet and VGG-16, in terms of saved communication time, speedup and accuracy.

\section{Data Parallelism of DNN}\label{sec:background} 
\subsection{Standard Data Parallelism}\label{subsec:plump-dp}
We first introduce the notations for the standard data parallelism with the popular parameter server architecture \cite{li2014communication,li2014scaling}. We denote a DNN model as $f(\mathbf{w})$, where $\mathbf{w}$ is the vector of the parameters. We assume that there are $K$ workers employed in the parallel architecture, and each worker, say the $k$-th worker, holds a local dataset $D_k=\{(x_{k,1},y_{k,1}),\dots, (x_{k,m_k},y_{k,m_k})\}$ with size $m_k$. We denote the local model and its update at the iteration $t$ on the worker $k$ as $\mathbf{w}_k^t$ and  $\delta_k^t$. Furthermore, if the local updates are communicated at iteration $t$, we denote the global model aggregated on the parameter server as $\bar{\mathbf{w}}^t$. The communication between the worker and the parameter server will be invoked after the worker conducts every $p$ iterations of local training on its local data. We call $p$ the communication frequency.

\begin{figure}[t]
	\centering
	\begin{minipage}{0.49\columnwidth}
		\centering
		\includegraphics[width=\columnwidth]{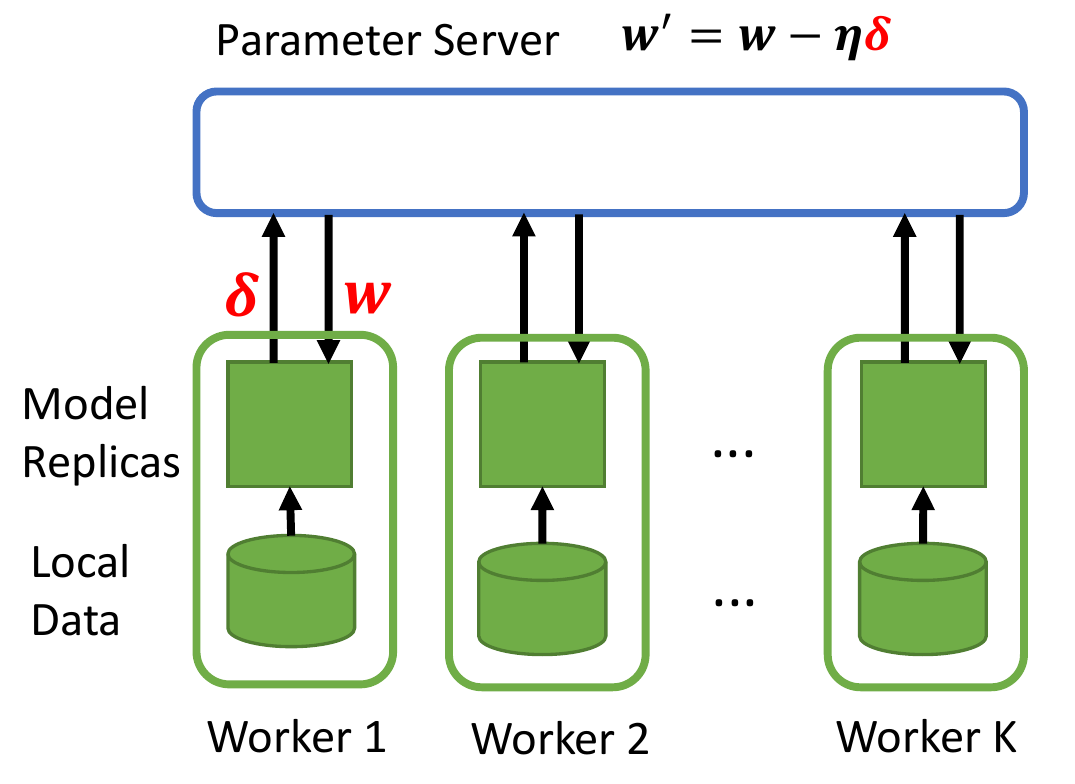}
		\caption{Plump-DP.}
		\label{fig:plump_dp}
	\end{minipage}
	\begin{minipage}{0.49\columnwidth}
		\centering
		\includegraphics[width=\columnwidth]{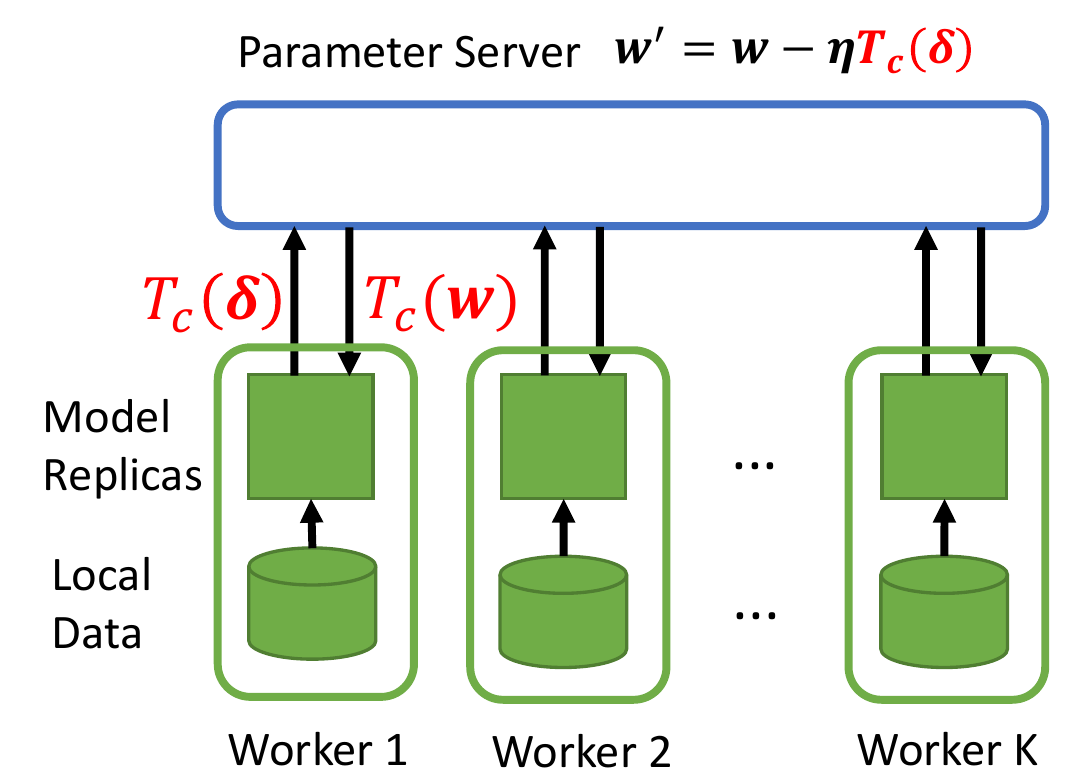}
		\caption{Slim-DP.}
		\label{fig:slim_dp}
	\end{minipage}
\end{figure}

As illustrated in Figure~\ref{fig:plump_dp}, a typical data parallelism for DNN~\cite{dean2012large,chen2016revisiting} iteratively implement three steps until the training procedure converges\footnote{In this paper, we focus on synchronous data parallelism, considering that synchronous data parallelism can achieve better convergence than asynchronous data parallelism~\cite{chen2016revisiting}. The algorithm and results can be easily generalized to the asynchronous mode as well.}.

\emph{1. Local training: } Each worker independently trains the local model based on its local data by stochastic gradient decent (SGD) or other stochastic algorithm. There will be no synchronization with the parameter server until every $p$ local training iterations. 

\emph{2. Push: } Each worker pushes the parameter updates of its local model to the parameter server, who will merge those updates with the current global model. 

\emph{3. Pull: } After the global model has been renewed by local updates, the local worker will pull the new snapshot of the global model from the parameter server, and set it as the starting point for the next round of local training.

Since this standard data parallelism approach always transfers updates and parameters of the whole DNN model, we call it \emph{Plump Data Parallelism}, abbreviated as \emph{Plump-DP} for ease of reference.

\subsection{Related Works}\label{subsec:related_work}
Many works improve the parallel training of DNN by designing new local training algorithms, new model aggregation methods, and new global model update rules. For example, to improve the local training, NG-SGD~\cite{povey2014parallel} implements an approximate and efficient algorithm for Natural Gradient SGD; large mini-batch methods~\cite{goyal2017accurate,you2017scaling} increase the learning rate and the mini-batch size to accelerate the convergence. To design new model aggregation methods, EASGD~\cite{zhang2015deep} adds an elastic force which takes the weighted combination of the local model and the global model as the new local model after the synchronization; EC-DNN~\cite{sun2017ensemble} uses the output-average instead of the parameter-average to aggregate local models in order to improve the convergence. To improve the global model update rules, BMUF~\cite{chen2016scalable} designs block-wise model-update filtering and utilizes the momentum of the global model to improve the speedup of Plump-DP. 

Since the parallel training for large DNN models suffers from the heavy communication cost, researchers and engineers improve the speed of the parallel training of DNN by reducing the communication cost. First, system-level technique can be used to alleviate the communication cost, in which the computation of one layer is made overlap with the communication of gradients of another layer~\cite{chen2016revisiting,goyal2017accurate}. Unfortunately, there is no system-level technique can perfectly hide all the communication time without loss of accuracy~\cite{chen2016revisiting}, and thus it is necessary to employ algorithm-level methods. Second, for NLP tasks, sampling method~\cite{xiao2017fast} only transfers the gradients of the parameters that corresponds to the most frequent words in the vocabulary in the RNN model. However, such method cannot be generalized to the parallel training of general DNN models. Furthermore, quantization method~\cite{seide2014,alistarh2016qsgd} quantizes each gradient to a small number of bits (less than 32 bits) during the communication, and restore the gradient to 32 bits again in the local training and the global model update. Considering that we communicate the computing information of the significant parameters while the quantization method uniformly sacrifices the precisions of different parameters, our method will perform better than the quantization method when the communication cost is similar. In the meanwhile, our method can be applied together with the quantization method, since the precision of each parameter in our method can be further reduced. 

\section{Slim-DP}\label{sec:slim_dp}
In this section, we propose a new data parallelism framework, called \emph{Slim Data Parallelism}, abbreviated as \emph{Slim-DP}, to address the challenge in terms of heavy communication cost of Plump-DP. 
\subsection{An Explore-Exploit Like Approach for Communication Set Determination}\label{subsec:core}
DNN is a highly redundant model, i.e., a considerable proportion of parameters of a DNN model is insignificant to the model performance. In the model compression task, such insignificant parameters in a well-trained model can be deleted via weight pruning without performance drop~\cite{han2015learning,lecun1990optimal,hassibi1993second}. Actually, such redundancy also exposes a great opportunity to reduce the communication cost in the parallel training of DNN, that we may only transfer the computing information of those significant parameters instead of all the parameters during the training process. For ease of reference, we call the subset of the parameters communicated during the parallel training \emph{communication set}.

However, it is improper to simply select the temporally significant parameters in the current global model as the communication set. The reason is that, different from removing the redundant parameters of a well-trained model in the model compression task, the selection of the communication set is conducted during the training process. The synchronization of the parameters in the selected communication set will influence the forthcoming optimization path. If we only communicate the information of the temporally significant parameters and give no opportunities to the communication of other parameters, we may miss the parameters that is insignificant now but have potential to become significant as the training process goes on if their computing information is communicated. 

Therefore, we design an Explore-Exploit framework to dynamically identify the communication set. In particular, this approach consists of two operators: 

\emph{1. Exploitation in parameter server side.} Parameter server employs an exploit-like \emph{core-selection} operator, denoted as $T_S$, aiming at picking the significant parameters. We call such set of significant parameters \emph{core} for ease of reference. In addition, the parameter values and updates in the core are denoted as $T_S(\bar{\mathbf{w}}^t)$ and $T_S(\delta_k^t)$ respectively. For better exploitation, the core-selection operator is invoked after each $q$ iterations of communication instead of at every communication because communicating the core for multiple times can help us learn the parameters in this set more sufficiently. 

\emph{2. Exploration in local worker side.} Local worker employ an explore-like \emph{exploration} operator, denoted as $T_{R}^k$, targeting randomly picking a set of parameters from those other than the core. We call such set of random parameters \emph{explorer} for ease of reference. In addition, the values and updates of the explorer are denoted as $T_{R}^k(\bar{\mathbf{w}}^t)$ and $T_{R}^k(\delta_k^t)$ respectively, where $k$ is the index of the local worker. For sufficient exploration, this operator is executed by each local worker for every communication.

As illustrated in Figure~\ref{fig:slim_dp}, in such Explore-Exploit like framework, when the communication happens, the local worker pushes the updates of the core and explorer, i.e., $T_C(\delta_k^t) \triangleq T_S(\delta_k^t)\cup T_{R}^k(\delta_k^t)$, to the parameter servers, and then pulls the latest parameters of the core and explorer, i.e.,  $T_C({\mathbf{w}}^t) \triangleq T_S({\mathbf{w}}^t)\cup T_{R}^k(\bar{\mathbf{w}}^t)$.  In this way of synchronization, the communication cost in the push step is $ T_C(\delta_k^t) $ and that in the pull step is $ T_C(\bar{\mathbf{w}}^t)$, which are quite smaller than synchronizing the entire global model as in Plump-DP, i.e., $\delta_k^t$ and $\bar{\mathbf{w}}^t$ as shown in Figure~\ref{fig:plump_dp}.

\subsection{Significance Evaluation}\label{subsec:significance}
 
 To measure the significance of the parameters, we consider two factors: the magnitude of this parameter and the magnitude of its gradient. On the one hand, if the magnitude of a parameter is close to zero, deleting this parameter will not significantly influence the model performance, which has also been already widely used in the model compression task~\cite{han2015learning}. On the other hand, the magnitude of the gradient of one parameter indicates the sensitivity of the loss function with respect to this parameter~\cite{lecun1990optimal,hassibi1993second}. If the magnitude of the gradient is large, the deletion of the parameter will cause significant change to the loss function even if the magnitude of the parameter is not such great.
 
 Therefore, we propose to measure the significance of the parameter $i$, $i\in\{1,\dots,n\}$, denoted as $S_i$, by the combination of the magnitude of this parameter $\lvert w_i\rvert$ and the magnitude of its gradient $\lvert g_i \rvert$. In our experiments, we employ a weighted sum format of the combination, i.e.,
 \begin{small}
 	\begin{align}
 	S_i = \lvert w_i \rvert + c \lvert g_i \rvert;\quad i\in\{1,\dots,n\}
 	\end{align}
 \end{small}where $c$ is a positive coefficient to adjust the magnitude of the parameter and its gradient to the same scale. We rank the parameters by their significance, and select the top several significant parameters as the core.

The benefit of communicating significant parameters is three-fold. First, since the significant parameters will be well-explored via the Explore-Exploit like approach mentioned in Section~\ref{subsec:core}, there will be almost no loss of accuracy. Second, by ignoring the redundant information, we can also achieve a similar effect like regularization~\cite{srivastava2014dropout,wan2013regularization}, which even benefits the convergence. Finally, unlike the set of random parameters that should be re-sampled frequently, the set of significance parameters will be relatively stable in recent training process, which will benefit the communication efficiency (see more details in Section~\ref{subsec:implenmentation}). 

\subsection{Algorithm Description}\label{subsec:alg}
Firstly, we introduce the inputs for Slim-DP as follows. 

1. The local data set $D_k$, where $k\in\{1,\dots,K\}$; 

2. The hyperparameter $\alpha\in[0,1]$ that controls the size of communication set in Slim-DP comparing to Plump-DP, i.e., $\alpha = \frac{|T_C(\delta_k^t)|}{|\delta_k^t|}=\frac{|T_C(\bar{\mathbf{w}}^t)|}{|\bar{\mathbf{w}}^t|}$; 

3. The hyperparameter $\beta\in[0,\alpha]$ that controls the size of the core. It is clear that $\beta = \frac{|T_S(\delta_k^t)|}{|\delta_k^t|}=\frac{|T_S(\bar{\mathbf{w}}^t)|}{|\bar{\mathbf{w}}^t|}$; 

4. The communication frequency $p$, i.e., the frequency that the local worker push $T_C(\delta_k^t)$ to or pull $T_C(\bar{\mathbf{w}}_k^t)$ from the parameter server. It is also the frequency that the exploration operation $T_{R}^k$ is executed by the local worker $k$; 

5. The core-selection frequency $q$, i.e., the frequency that the core-selection operator $T_S$ is executed by the parameter server, in which a new core is selected according to the significance of the parameters in the current global model $\bar{\mathbf{w}}^t$.

Secondly, we introduce the Slim-DP algorithm (described in Algorithm~\ref{alg:subgraph}), which consists of the following six steps. 
	\begin{small}
		\begin{algorithm}
			\caption{Slim-DP ($D_k,\alpha,\beta,p,q$)}
			\label{alg:subgraph}
			\textbf{Local Worker $\mathbf{k}$:}\\
			$\mathbf{w}_k^0\gets$Pull($\bar{\mathbf{w}}^0$)\;
			\While{$f(\bar{\mathbf{w}}^t;x)$ dose not converge}{			
				$\delta_k^t\gets$ LocalTrain($\mathbf{w}_k^t,D_k,p$)\;
				$T_{R}^k\gets$ Exploration($\delta_k^t\backslash T_S(\delta_k^t),\alpha-\beta$)\;
				\If{$(t+1)\%q==0$}{
					Push$(\delta_k^t)$\;
				}
				\Else{Push$(T_C(\delta_k^t))$\;
				}			 
				$\tilde{\mathbf{w}}_k^t\gets$ Pull$(T_C(\bar{\mathbf{w}}^t))$\; 
				$\mathbf{w}_k^t \gets$ Merge($\mathbf{w}_k^t,\tilde{\mathbf{w}}_k^t$)\;
			}	
			\textbf{Parameter Server:}\\
			Randomly initialize $\bar{\mathbf{w}}^0$ and set $t=0$\;	
					
			\While{$f(\bar{\mathbf{w}}^t;x)$ does not converge}{
				$\bar{\mathbf{w}}^{t+1}\gets$
				Update$(T_C(\delta_k^t))$\;
				$t\gets t+1$\;
				\If{$t\%q == 0$}{
					$T_S\gets$ 
					Core-Selection($\bar{\mathbf{w}}^t,\delta_k^t, \beta$)\;	
				}
			}
			\Return $\bar{\mathbf{w}}^t$.\\
		\end{algorithm}
	\end{small}
	
\emph{1. LocalTrain(${\mathbf{w}}_k^t,D_k,p$): } Parameters of the local model ${\mathbf{w}}_k^t$ are updated by minimizing the cross entropy loss using SGD on its local dataset $D_k$, i.e., ${\mathbf{w}}_k^{t+1}={\mathbf{w}}_k^t-\eta\nabla({\mathbf{w}}_k^t,D_k)$, where $\eta$ is the learning rate and $\nabla(\mathbf{w}_k^t,D_k)$ is the gradients of the local model $\mathbf{w}_k^t$ on one mini batch of the local dataset $D_k$. Such update lasts for $p$ mini-batches before the communication with the parameter server. We accumulate the model updates over $p$ mini-batches and denote the result as $\delta_k^t$.

\emph{2. Exploration($\delta_k^t\backslash T_S(\delta_k^t)$, $\alpha-\beta$): } Each local worker samples $\alpha-\beta$ of updates from the parameters outsides the core, i.e., $\delta_k^t\backslash T_S(\delta_k^t)$. Note that only local worker $k$ will push the updates corresponding to this random set of parameters. This random set will be resampled before each iteration of communication, i.e., operation $T_{R}^k$ will be redefined before each iteration of communication.

\emph{3. Push($T_C(\delta_k^t)$)} and \emph{ Update($T_C(\delta_k^t)$): } At the local worker side, we execute Push($T_C(\delta_k^t)$), i.e., push the subset of local updates $T_C(\delta_k^t)$ to the parameter sever. Note that if it is the last communication before the core is reselected (i.e., when $(t+1)\%q == 0$), we push all the updates (i.e., $\delta_k^t$) to the parameter server in order to prepare for the computation of the significance of the parameters. At the parameter sever side, we execute Update($T_C(\delta_k^t)$), i.e., add $T_C(\delta_k^t)$ to the corresponding global parameters on the parameter server, i.e., $\bar{\mathbf{w}}^{t+1}= \bar{\mathbf{w}}^{t}-\eta^\prime T_C(\delta_k^t)$.

\emph{4. Pull($T_C(\bar{\mathbf{w}}^t)$): }We pull the subset of the parameters in the global model $T_C(\bar{\mathbf{w}}^t)$ from the parameter server, including both the core and the explorer. 

\emph{5. Merge($\mathbf{w}_k^t,\tilde{\mathbf{w}}_k^t$): }The parameters $\tilde{\mathbf{w}}_k^t$, which is pulled from the parameter server according to operation $T_C$, will be merged with the current local parameters $\mathbf{w}_k^t$ to produce a new local model. This new model will be set as the starting point of the next round of local training.

\emph{6. Core-Selection($\bar{\mathbf{w}}^t,\delta_k^t, \beta$): }At the parameter server side, the core is reselected according to the significance of parameters, i.e., operation $T_S$ is redefined. Here the significance is related to the magnitude of parameters in the global model $\bar{\mathbf{w}}^t$ and gradients $\delta_k^t$, as discussed in Section~\ref{subsec:significance}. Note that we use a old version of gradients for the computation of the significance to save time for extra backward propagation and communication. 

All the steps are executed iteratively until the training converges. After every $q$ rounds from the first to the fifth step, Slim-DP will perform the sixth step to renew the core. 
\subsection{Discussions}\label{subsec:alg_discuss}
We make the following discussions for Slim-DP.

\noindent\textbf{Trade-off between Accuracy and Speed.}
The communication cost $\alpha$ trades-off the accuracy and the speed. On the one hand, larger $\alpha$ indicates that Slim-DP can communicate more parameters, including both the core and the explorer, which will result in better accuracy. The reason is that a larger core ensures the coverage of sufficient number of significant parameters, and a larger set of random explored parameters results in sufficient exploration outside the core. On the other hand, larger $\alpha$ also implies more communication cost, which slows down the training.  
 
\noindent\textbf{Trade-off between Exploration and Exploitation.}
For a fixed communication cost $\alpha$, the value of $\beta$ trades-off the exploration and the exploitation.
On the one hand, when $\beta$ yields a greater value or even $\beta=\alpha$, there is no sufficient exploration of parameters other than those in the core during the training, which may cause that some parameters that could have been significant cannot receive sufficient learning. On the other hand, when $\beta$ is very small or even $\beta=0$, the selected core cannot cover enough significant parameters, which will hurt the performance as well.

\noindent\textbf{Relationship with Dropout/DropConnect.}
The motivation of the two methods are different. Slim-DP aims at designing a data parallelism approach to accelerate the training of big DNN models by reducing communication cost, while Dropout/ Dropconnect~\cite{srivastava2014dropout,wan2013regularization} is a trick in the sequential training to avoid overfitting. Moreover, core-selection is quite indispensable for Slim-DP, while Dropout/DropConnect simply applies pure random sampling. 

\subsection{Communication and Time Efficiency}\label{subsec:implenmentation}
In this subsection, we analyze the communication cost and  the extra time in Slim-DP, and make comparison with Plump-DP.

\noindent\textbf{Communication Efficiency.}
In Slim-DP, since we transfer a subset instead of the whole set of the parameters, we should make the receiver know which parameters are transferred. Therefore, the information transfered between local workers and the parameter server should be represented as $\langle$key, value$\rangle$ pairs, where the key is the index of the parameter and the value is the corresponding parameter or its update. Thus, the real amount of transferred information has to double the size of the communicated parameters or updates. This contradicts with our motivation to reduce communication cost.

To tackle this challenge, for the information about the core, we use key catching filter~\cite{li2014communication} to transfer it because the core is not frequently renewed during the training and thus the keys for the core can keep unchanged for a long period. In the key catching filter, both the sender and receiver have cached the keys, and the sender then only needs to send the values with a signature of the keys. Therefore, the real amount of transferred information for the core is reduced to the same size of the core, i.e., $\beta n$, where $n$ is the dimension of the parameters. For the information about the explorer, we still transfer it by the $\langle$key, value$\rangle$ pair, and thus the real amount of transferred information for the explorer is $2(\alpha-\beta)n$.

Overall, in Slim-DP, the real amount of total transferred information is $(2\alpha-\beta) n$. In Plump-DP, the real communication amount equals to the dimension of the parameters, i.e., $n$, since we need to transfer the whole set of parameters.

\noindent\textbf{Time Efficiency.}
Compared to Plump-DP, Slim-DP may bring in two kinds of potential extra time. The first kind of potential extra time is the time to generate the exploration operator and the core-selection operator.
The generation of the exploration operator can be overlapped with the gradient computation at the local worker side, while the generation of the core-selection operator can be overlapped with gradient update at the parameter server side. Therefore, generating the exploration operator and core-selection operator will not bring in extra time. 

The second kind of potential extra time is the time to extract corresponding parameters/updates from the whole set according to the exploration and the core-selection operator. In the worst case, such extraction is done by scanning the whole set of the parameters/updates, whose time cost is proportional to the dimension of the parameters $n$. In practical implementation, multi-thread scanning can be easily leveraged to ensure limited such time cost. 

Overall, Slim-DP will bring a very small amount of extra time, which is less than $\mathcal{O}(n)$.

\section{Experiments}\label{sec:exp}
\begin{table*}[t]
	\centering
	\caption{Computational and communication time (hours) for each local worker to process 10k mini-batches of data.}
	\label{tab:time}
	\begin{small}
	\begin{tabular}{|c|c|c|c|c|c|c|c|c|}
		\hline
		& \multicolumn{4}{c|}{K=4}                                                      & \multicolumn{4}{c|}{K=8}                                                 \\ \hline
		& \multicolumn{2}{c|}{GoogLeNet}        & \multicolumn{2}{c|}{VGG-16}           & \multicolumn{2}{c|}{GoogLeNet}        & \multicolumn{2}{c|}{VGG-16}      \\ \hline
		& T$_{comp}$              & T$_{comm}$      & T$_{comp}$              & T$_{comm}$      & T$_{comp}$              & T$_{comm}$      & T$_{comp}$              & T$_{comm}$ \\ \hline
		Plump-DP & \multirow{3}{*}{2.28} & 0.40          & \multirow{3}{*}{7.83} & 4.09          & \multirow{3}{*}{2.32} & 0.57          & \multirow{3}{*}{7.82} & 5.51     \\ \cline{1-1} \cline{3-3} \cline{5-5} \cline{7-7} \cline{9-9} 
		Quant-DP &                       & 0.20          &                       & 1.47          &                       & 0.29          &                       & 1.93     \\ \cline{1-1} \cline{3-3} \cline{5-5} \cline{7-7} \cline{9-9} 
		Slim-DP  &                       & \textbf{0.18} &                       & \textbf{1.18} &                       & \textbf{0.25} &                       & \textbf{1.65}     \\ \hline
	\end{tabular}
	\end{small}
\end{table*}

\begin{table*}[]
	\centering
	\caption{Top-5 test accuracy (\%) and speedup.}
	\label{tab:acc}
	\begin{small}
	\begin{tabular}{|c|c|c|c|c|c|c|c|c|c|c|c|c|}
		\hline
		& \multicolumn{6}{c|}{K=4}                                                             & \multicolumn{6}{c|}{K=8}                                                               \\ \hline
		& \multicolumn{3}{c|}{GoogLeNet}                 & \multicolumn{3}{c|}{VGG-16}         & \multicolumn{3}{c|}{GoogLeNet}                 & \multicolumn{3}{c|}{VGG-16}           \\ \hline
		& Acc            & Speed$_{d}$   & Speed$_{a}$   & Acc   & Speed$_{d}$   & Speed$_{a}$ & Acc            & Speed$_{d}$   & Speed$_{a}$   & Acc   & Speed$_{d}$   & Speed$_{a}$   \\ \hline
		Plump-DP & 88.06          & 1             & 1             & 86.53 & 1             & 1           & 88.03          & 1             & 1             & 86.48 & 1             & 1             \\ \hline
		Quant-DP & 88.02          & 1.08          & 1.08          & 86.55 & 1.28          & 1.30        & 88.08          & 1.11          & 1.16          & 86.53  & 1.37          & 1.39          \\ \hline
		Slim-DP  & \textbf{88.29} & \textbf{1.09} & \textbf{1.16} & \textbf{87.03} & \textbf{1.32} & \textbf{1.45}        & \textbf{88.26} & \textbf{1.13} & \textbf{1.23} & \textbf{86.91} & \textbf{1.41} & \textbf{1.51} \\ \hline
	\end{tabular}
	\end{small}
\end{table*}
\subsection{Experimental Settings}\label{subsec:exp_setting}
\noindent\textbf{Platform.}
Our experiments are conducted on a GPU cluster interconnected with an InfiniBand network, each machine of which is equipped with two NVIDIA's K20 GPU processors. One GPU processor corresponds to one local worker.

\noindent\textbf{Data.} 
We conduct experiments on ImageNet (ILSVRC 2015 Classification Challenge)~\cite{ILSVRC15}. In our experiments, each image is normalized by subtracting the per-pixel mean computed over the whole training set, and cropped to the size of 224$\times$224. In addition, no data augmentation is used during the training. 

\noindent\textbf{Model.} 
We employ two models, i.e., VGG-16~\cite{simonyan2014very} and GoogLeNet~\cite{szegedy2015going}. VGG-16 is a 16-layer convolutional neural network with about 140M parameters and GoogLeNet is a 22-layer convolutional neural network with about 13M parameters. All the hyperparameters of the models are set the same as that in the Caffe~\cite{jia2014caffe} model zoo. 

\noindent\textbf{Parallel Setting.} 
We explore the number of workers $K \in \{4, 8\}$. Local workers communicate with the parameter server after the updates for every mini-batch, i.e., $p=1$. We use DMTK framework\footnote{\url{https://github.com/Microsoft/multiverso}} to implement the related operations of the parameter server.

\subsection{Compared Methods}
We compare performance of the following three methods.
\begin{itemize}
\item\textbf{Plump-DP} denotes the standard data parallelism framework that transfers the updates and parameters of the whole global model~\cite{dean2012large,chen2016revisiting}. 

\item\textbf{Quant-DP} denotes the method that reduces the communication cost by quantizing each gradient to a small number of bits (less than 32 bits) during the communication. There are two kinds of such quantization method, i.e. 1-bit SGD~\cite{seide2014} and random quantization SGD~\cite{alistarh2016qsgd}. In our experiments, we implement the latter one since it yields better performance by introducing randomization. And, we use the same hyper-parameters as in~\cite{alistarh2016qsgd}, i.e., we employ the 8-bit version and set the bucket size as 512.

\item\textbf{Slim-DP} refers to the data parallelism framework proposed in this paper, which reduces the communication cost by transferring the significant parameters (i.e., the core) together with a random explored set of other parameters (i.e., the explorer). Without special statement, we set $\alpha=0.2$ and $\beta=0.1$ for VGG-16, and $\alpha=0.3$ and $\beta=0.15$ for GoogLeNet. For the core-selection frequency $q$, we set $q=20k$ mini-batches for VGG-16 and $q=50k$ mini-batches for GoogleNet.
\end{itemize}



\subsection{Experimental Results}\label{subsec:exp_result}
\begin{figure*}[t]
	\centering
	\subfloat[GoogLeNet, $K=4$]
	{
		\includegraphics[width=0.5\columnwidth]{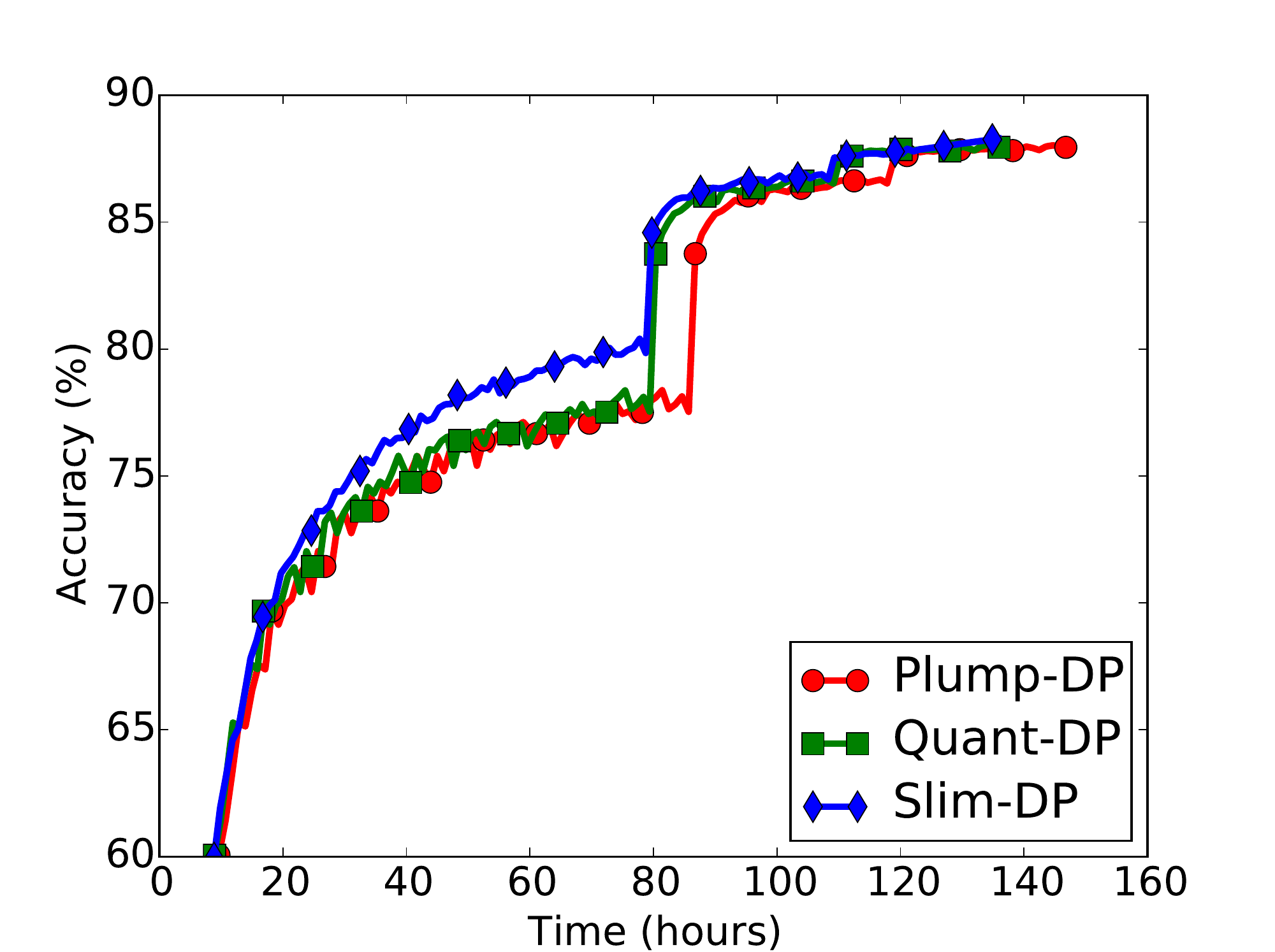}
		\label{subfig:basic_gl_4}
	}
	\subfloat[VGG-16, $K=4$]
	{
		\includegraphics[width=0.5\columnwidth]{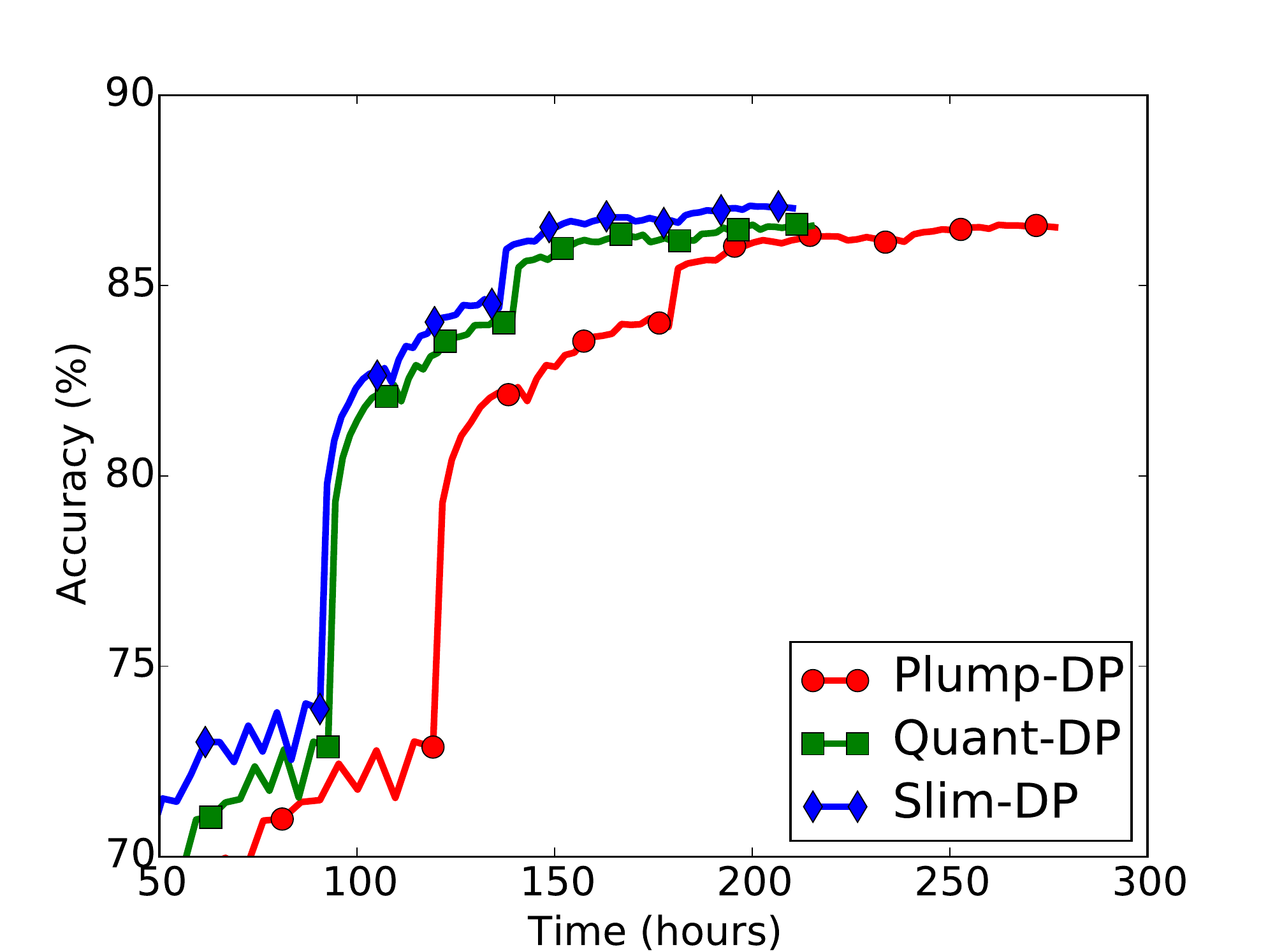}
		\label{subfig:basic_vg_4}
	}
	\subfloat[GoogLeNet, $K=8$]
	{
		\includegraphics[width=0.5\columnwidth]{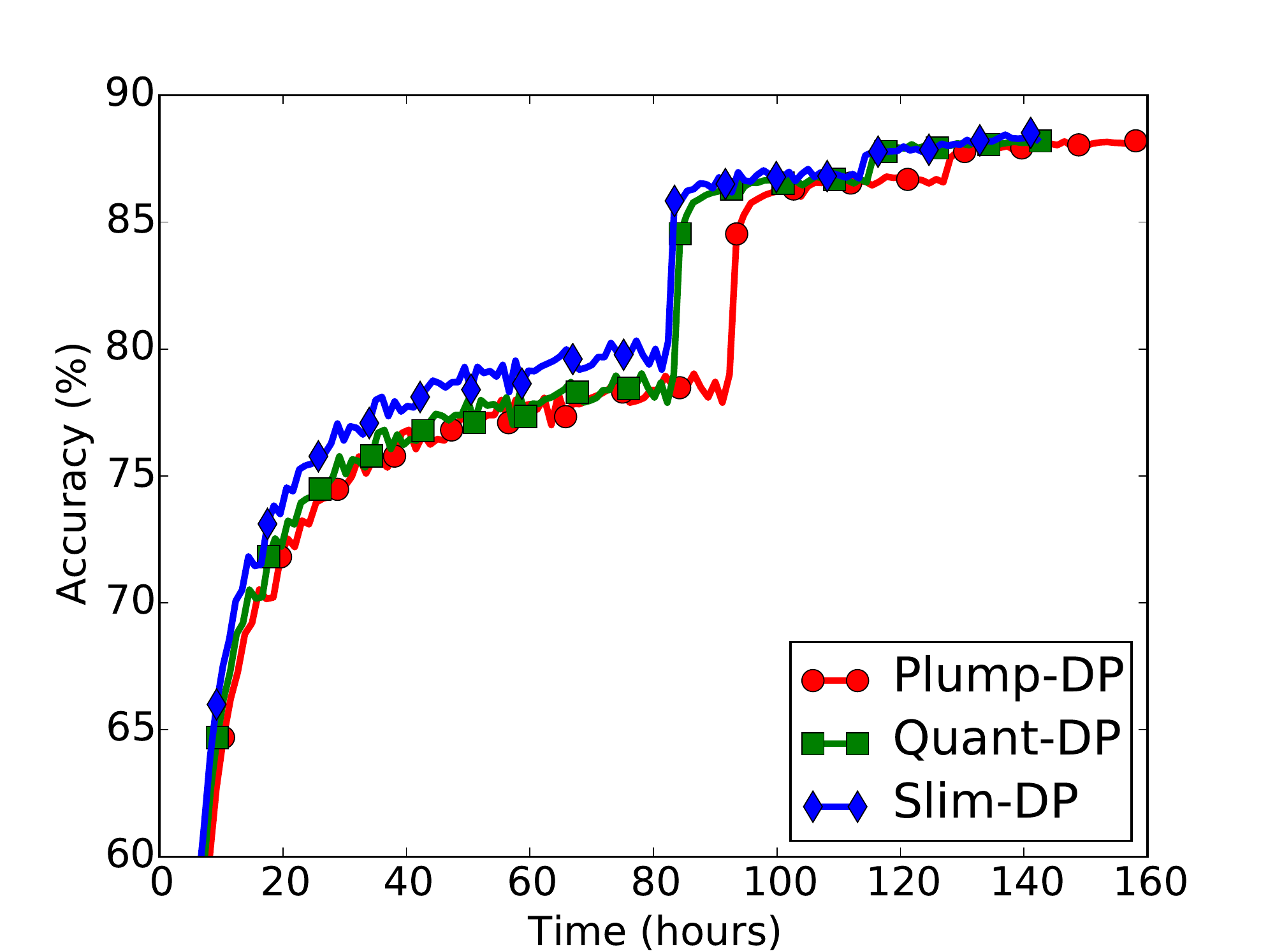}
		\label{subfig:basic_gl_8}
	}
	\subfloat[VGG-16, $K=8$]
	{
		\includegraphics[width=0.5\columnwidth]{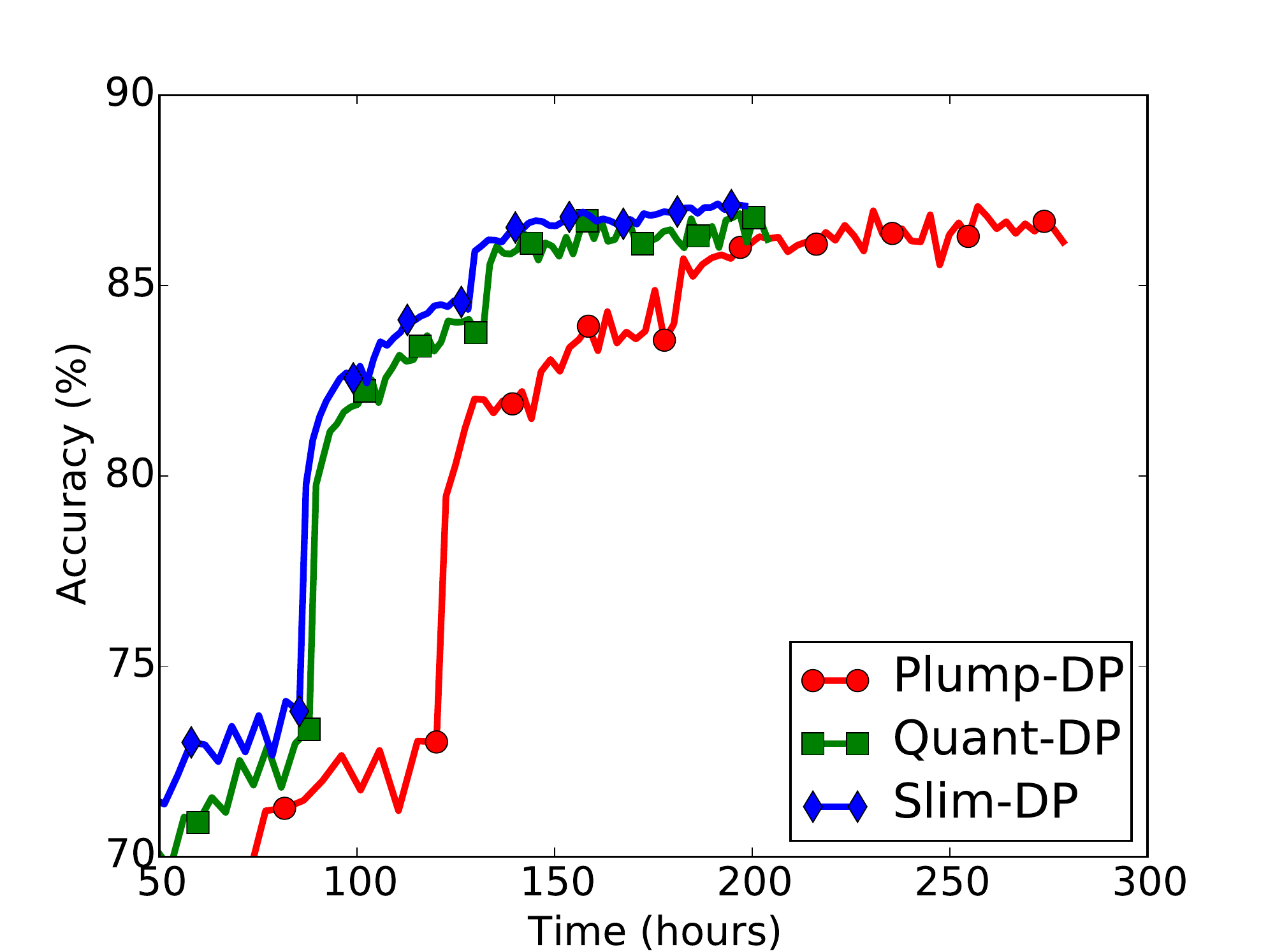}
		\label{subfig:basic_vg_8}
	}
	\caption{Top-5 Test Accuracy w.r.t. Time}
	\label{fig:basic_acc}
\end{figure*}
\subsubsection{Communication Time vs. Computational Time.}
We first compare the communication time and computational time of all the methods. To this end, we count the communication time and computational time for each local worker to process 10k mini-batches of data. For Quant-SGD, the extra decoding and encoding time has been counted into the communication time. For Slim-DP, the extra time to extract parameters/updates according to exploration and core-selection (see Section~\ref{subsec:implenmentation} for the analysis of the time efficiency) has been counted into the communication time.

Table~\ref{tab:time} shows the communication time (denoted as T$_{comm}$) and computational time (denoted as T$_{comp}$) of each method on both GoogleNet and VGG-16. From this table, we can observe that for Plump-DP, communication time is indeed non-negligible for both GoogleNet and VGG-16. Specifically, the communication time of Plump-DP takes about 15\% and 34\% of the overall time for GoogLeNet and VGG-16 when $K=4$, and it even takes more percentage when $K$ is increased to $8$. Furthermore, Slim-DP can reduce the communication time with a better efficiency than Quant-DP. For example, when $K=4$, Slim-DP save about 55\% and 70\% of the communication time for GoogLeNet and VGG-16 respectively, while Quant-DP save about 50\% and 65\% of the communication time respectively. The observations of $K=8$ are similar. 

\subsubsection{Speedup.}
We compare two kinds of speedup of Quant-DP and Slim-DP over Plump-DP. The first type is the speed increase for each local worker to process the same number of mini-batches of the data. We calculate such speedup from Table~\ref{tab:time} for all the methods and denote it as \emph{Speed}$_d$ in Table~\ref{tab:acc}. The second type is the speed increase of how fast to achieve the same accuracy as Plump-DP when Plump-DP converges. We calculate such speedup from Figure~\ref{fig:basic_acc} and denote it as \emph{Speed}$_a$ in Table~\ref{tab:acc}. For better demonstration, we normalize the speed of Slim-DP, Quant-DP, and Plump-DP by dividing each of them by that of Plump-DP.

For Speed$_d$, we have following observations. First, Slim-DP uses less time than Plump-DP to process the same number of mini-batches of the data. For example, when $K=8$, even for computation-intensive model such as GoogLeNet, Slim-DP run about $1.13$ times faster than Plump-DP, while for communication-intensive model such as VGG-16, the speedup is improved to $1.41$. In addition, Slim-DP also runs faster than Quant-DP. For example, for VGG-16, when $K=4$, the speedup of Quant-DP is $1.28$ while the speedup of Slim-DP is $1.32$.

For Speed$_a$, we have following observations. First, Slim-DP runs much faster than Plump-DP to achieve the same accuracy. For example, when $K=4$, Slim-DP can run $1.16$ and $1.45$ times faster than Plump-DP for GoogLeNet and VGG-16 respectively. Second, Slim-DP achieves better speedup than Quant-DP. For example, for GoogLeNet, when $K=4$, the speedup of Quant-DP is $1.08$ while the speedup of Slim-DP is $1.16$. Finally, Speed$_a$ is higher than Speed$_d$ for Slim-DP. For example, for $K=4$ and GoogLeNet, Speed$_d$ is $1.09$ while Speed$_a$ is $1.16$ for Slim-DP. This observation is consistent with our discussion in Section~\ref{subsec:significance}, that by ignoring insignificant information, Slim-DP can achieve a similar effect as regularization, which benefit the convergence.

\subsubsection{Accuracy.}
We compare the accuracy of all the methods when the models are trained to the convergence. From Table~\ref{tab:acc} (where the accuracy is denoted as \emph{Acc}) and Figure~\ref{fig:basic_acc}, we observe that Slim-DP even achieves better accuracy than Plump-DP and Quant-DP. The accuracy improvement of Slim-DP over Plump-DP and Quant-DP is about $0.2\%$ and $0.5\%$ for GoogLeNet and VGG-16 respectively. These observations verified our discussion on the benefits of considering parameters' significance when reducing communication cost in Section~\ref{subsec:significance}.

\subsubsection{Trade-off between Exploration and Exploitation.}
To investigate the effects of exploration and exploitation, we fix the size of the communication set (i.e., $\alpha$) and vary the size of the core (i.e., $\beta$). We set $\alpha=0.3$, and compare the performance of Slim-DP when $\beta=0$ (no exploitation), $\beta = 0.15$ (the one used in the former experiments), and $\beta=0.3$ (no exploration). For ease of reference, we denote Slim-DP with constraint $\alpha$ on the communication set and the ratio $\beta$ that controls the size of the core as Slim-DP ($\alpha$, $\beta$). 

\begin{figure}[t]
	\centering
	\subfloat[]{
		\includegraphics[width=0.5\columnwidth]{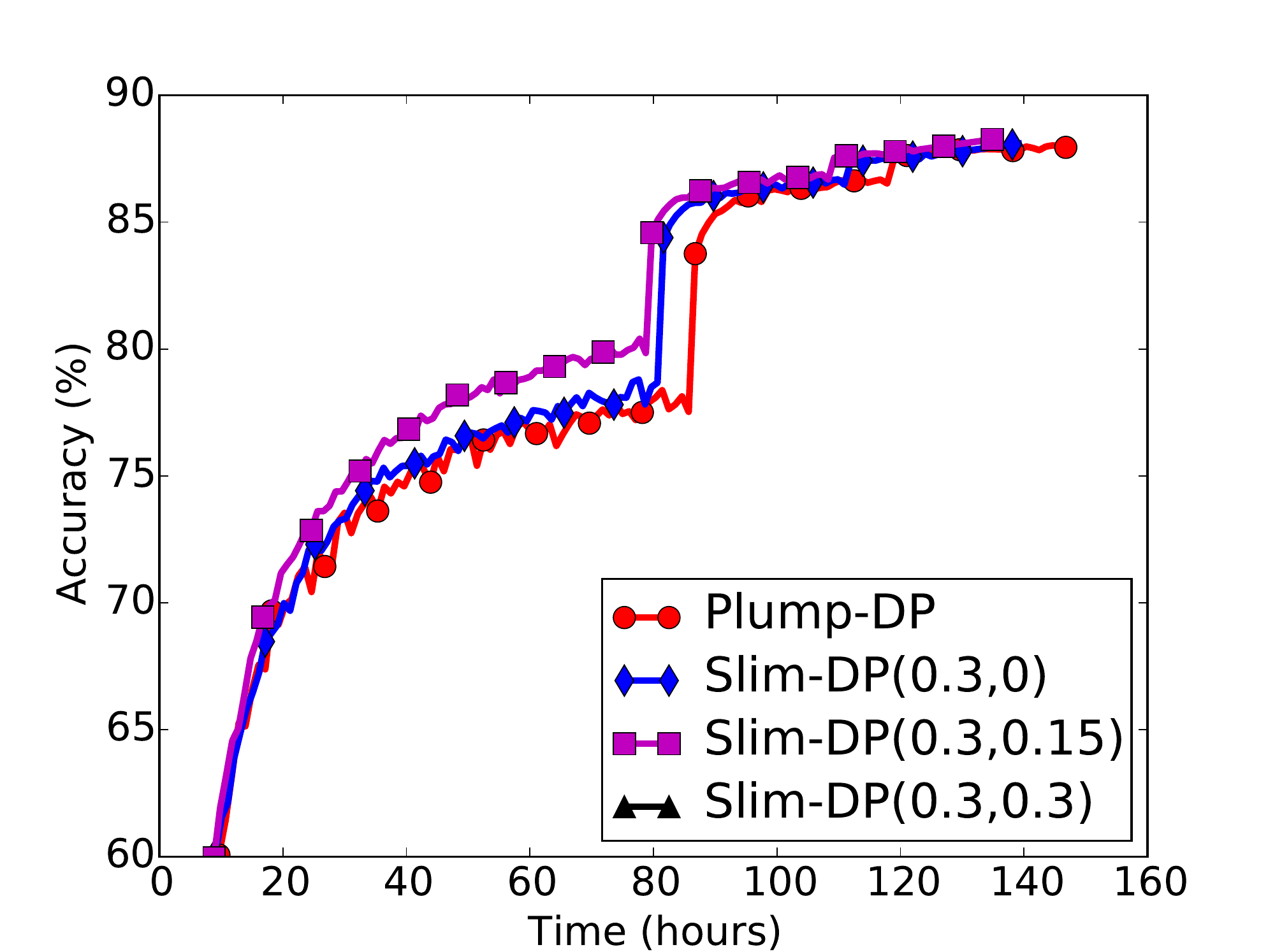}
		\label{subfig:alpha_gl_4}
	}
	\subfloat[]{
		\includegraphics[width=0.5\columnwidth]{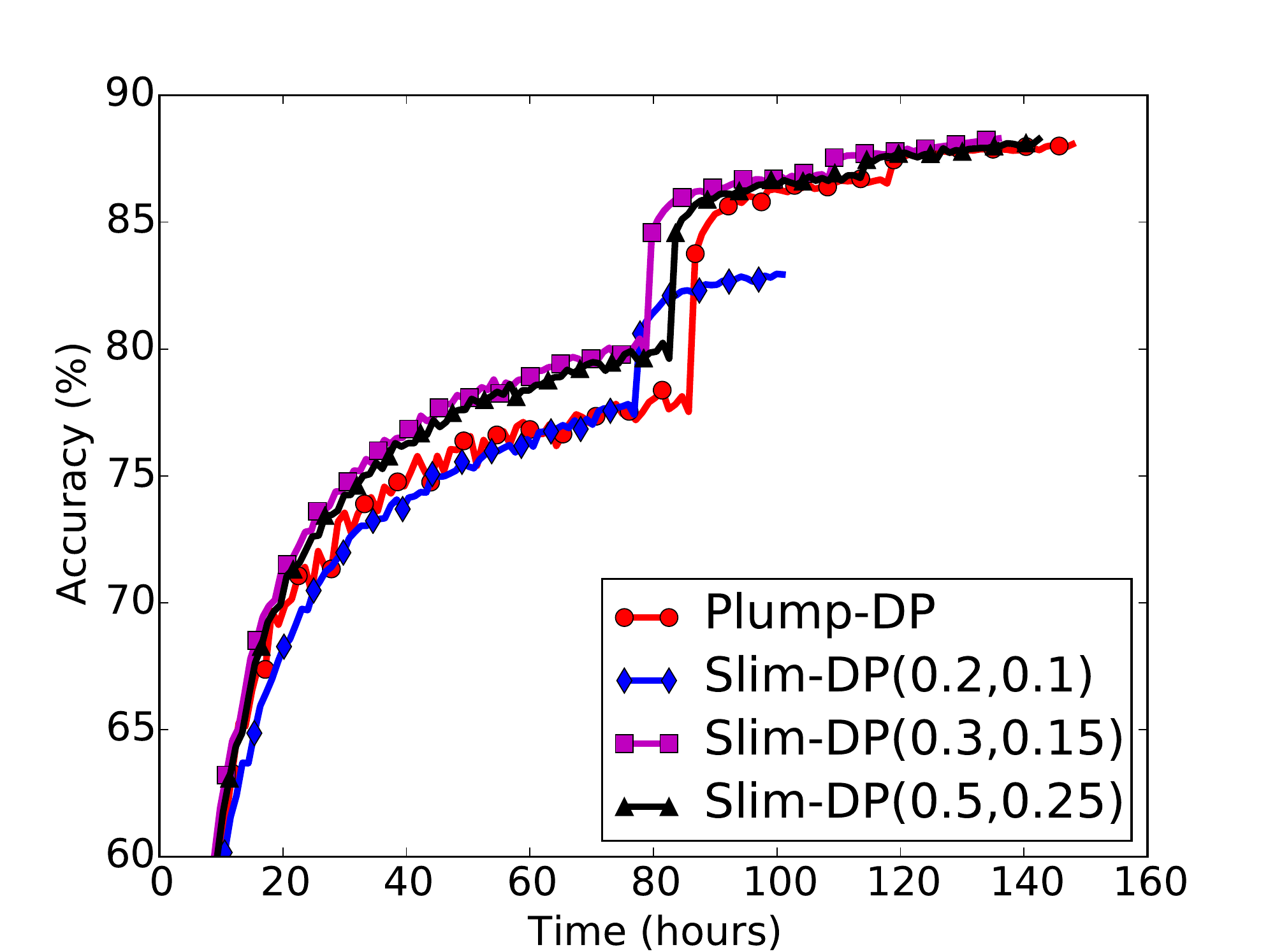}
		\label{subfig:beta_gl_4}
	}
	\caption{Trade-off in Slim-DP. (a) Exploration and Exploitation Trade-off. (b) Accuracy and Speedup Trade-off.}
	\label{fig:trade-off}
\end{figure}

Figure~\ref{subfig:alpha_gl_4} shows the test accuracy curves w.r.t. the overall time. Note that we take GoogLeNet and $K=4$ as an example and the observations on VGG-16 and $K=8$ are similar. From this figure, we can observe that Slim-DP (0.3,0.15), which considers both exploration and exploitation, achieves best performance, indicating that both exploration and exploitation are indispensable for the success of Slim-DP. When there is no exploitation, i.e., Slim-DP (0.3,0), Slim-DP equals to DropConnect. We observe that Slim-DP in such case slightly improve the performance of Plump-DP in terms of accuracy as a regularization method. When there is no exploration, i.e., Slim-DP (0.3,0.3), Slim-DP fails to converge and thus we do not show it in the figure. 

\subsubsection{Trade-off between Accuracy and Speedup.}
We plot the test accuracy curve w.r.t. the overall time for Slim-DP (0.2, 0.1), Slim-DP (0.3, 0.15) (the one that we used in the former experiments) and Slim-DP (0.5, 0.25) in Figure~\ref{subfig:beta_gl_4}. We take GoogLeNet and $K=4$ as an example and the observations on VGG-16 and $K=8$ are similar. We fix the ratio of the core w.r.t.the size of communicated parameters (i.e., fix $\beta/\alpha$) to avoid extra influence introduced by the trade-off between exploration and exploitation.

From Figure~\ref{subfig:beta_gl_4}, we observe that Slim-DP (0.3,0.15) achieves both best speedup and accuracy. For Slim-DP (0.2,0.1), it cannot achieve the same accuracy as Plump-DP since it communicates too few parameters and cannot cover enough significant parameters. For Slim-DP (0.5,0.25), although it achieves the similar accuracy as Slim-DP (0.3, 0.15), it does not achieve the similar speedup as Slim-DP (0.3, 0.15) because it transfers more parameters.

\section{Conclusion and Future Work}\label{sec:conclusion}
In this paper, we propose a novel approach, called Slim-DP, to reduce the communication cost of traditional data parallelism approach, called Plump-DP. Specifically, we only transfer a subset of the DNN model for communication by exploiting the significant parameters in the latest snapshot of the global model together with a random explored set of other parameters. Experimental results demonstrate that Slim-DP achieves better speedup than Plump-DP and its communication-efficient version (i.e., Quant-DP) without loss of accuracy. In the future, we will design model-specific significance measure for different types of models to further improve the performance.

\bibliographystyle{aaai}
\bibliography{dp}

\end{document}